\documentclass[prb,amsmath,amssymb,showpacs,twocolumn]{revtex4}

\usepackage[
  paperwidth=210mm,paperheight=297mm,
  centering,hmargin=1.7cm,vmargin=2.2cm]{geometry}

\pdfoutput=1
\usepackage{graphicx}
\usepackage{dcolumn}
\usepackage{bm}
\usepackage{hyperref}
\usepackage{amsmath}

\begin{document}

\title{Distributional exact diagonalization formalism for quantum impurity models}
\author{Mats Granath$^1$, Hugo U.~R. Strand$^1$}
\affiliation{$^1$Department of Physics, University of Gothenburg,
SE-41296 Gothenburg, Sweden}

\date{\today}

\begin{abstract}
We develop a method for calculating the self-energy of a quantum impurity coupled to a continuous bath by stochastically generating a distribution of finite Anderson models that are solved by exact diagonalization, using the noninteracting local spectral function as a probability distribution for the sampling.
The method enables calculation of the full analytic self-energy and single-particle Green's function in the complex frequency plane, without analytic continuation, and can be used for both finite and zero temperature at arbitrary fillings. Results are in good agreement with imaginary frequency data from continuous-time quantum Monte Carlo calculations for the single impurity Anderson model and the two-orbital Hubbard model within dynamical mean field theory (DMFT) as well as real frequency data for self energy of the single band Hubbard model within DMFT using numerical renormalization group. 
The method should be applicable to a wide range of quantum impurity models and particularly useful when high-precision real frequency results are sought.

\end{abstract}


\maketitle
The topic of quantum impurity models is a cornerstone in the study of strongly correlated electrons. The standard model is the Anderson impurity model which was first developed to describe dilute magnetic ions in metals.~It contains the most classic manifestation of strong correlations, the Kondo effect.~A modern application of quantum impurity models has emerged in the study of strongly interacting lattice models using dynamical mean field theory (DMFT), \cite{Georges:1996aa, Kotliar:2006aa} where the periodic system is mapped to an quantum impurity model coupled to a non-interacting bath, by assuming that correlations described by the self-energy, are fully dynamic but local. Although this assumption is exact only in the limit of infinite dimensions, DMFT has proven of great value in the study of finite dimensional correlated systems. Despite of providing a drastic simplification compared to the full momentum dependent lattice problem, solving the quantum impurity problem is still a major challenge.
\\ \indent
With the advent of high resolution angle resolved photoemission spectroscopy as one of the main probes of strongly correlated systems there is an increasing demand for calculations of the single particle spectral function, using for example DMFT.  In this context the most versatile and widely used impurity solver is the continuous time quantum Monte Carlo (CT-QMC) method.\cite{Gull:2011lr}
The sampling of the thermal Green's function in imaginary time or frequency is very efficient but, statistical errors grow with decreased temperature. Calculating the spectral function from the CT-QMC results requires a numerical analytic continuation to the real frequency axis. Here the statistical errors prevents the use of Pad\'e based methods \cite{Vidberg:1977aa, Granath:2011lr} while the maximum entropy method (MAXENT) \cite{Gubernatis:1991qy, Jarrell:1996fj} can be used to find a smoothed out spectral function. To be able to capture the detailed spectral function it is clearly preferable to work directly with real frequencies or time.
The numerical renormalisation group formalism (NRG) \cite{Bulla:2008kx} does exactly this and has been used to extract fine details such as small kinks in the quasiparticle dispersion caused by correlation effects. \cite{Byczuk:2007kx, Grete:2011yq} Nevertheless, the method is very demanding for models with more than one impurity site.
\\ \indent
Another very versatile tool for DMFT calculations is the so called exact diagonalization (ED) formalism \cite{Caffarel:1994aa} in which the quantum impurity problem is represented by an exactly solvable finite Anderson impurity model with a small set of non-interacting bath levels. The fit of the few bath levels to the continuum non-interacting impurity Green's function can be surprisingly accurate on the imaginary frequency axis and it is possible to calculate the corresponding self-energy to very good accuracy.~\cite{Liebsch:2011vn} However, if the impurity self-energy is evaluated along the real frequency axis the finite-size effects of the few-level Anderson model become apparent, giving a discrete set of poles (see eg.~Fig.~18 in Ref.~\onlinecite{Georges:1996aa}), while the self-energy of the impurity coupled to a continuous bath should be continuous. 
\\ \indent
In this paper we present a formalism for calculating the full analytic (real and imaginary frequency) self energy of a quantum impurity model by using a stochastic sampling of the non-interacting impurity spectral function instead of the Matsubara fit of the ED formalism. Each sample is a small $n$-level Anderson impurity model for which the self energy can be calculated by exact diagonalization. By sampling over the  impurity Green's function the full phase space can be explored and a, for practical purposes, continuous self energy is formed as a sample average.
\\ \indent
In terms of applications we believe that the method is as versatile as the ED formalism but with greater physical relevance. Comparisons with CT-QMC shows very good agreement on the Matsubara frequencies as well as with NRG real frequency results for the self energy. Compared to CT-QMC a major strength of the method is that real frequency results are calculated without any analytic continuation and that the method is most efficient for zero temperature. 
Because a large stochastic sampling is required for accurate results the calculations can be numerically demanding. However, the samples can be generated and addressed independently making it ideal for parallel computing. In this paper which introduces the method we will focus on the single-impurity Anderson model with a semicircular bare density of states, and the corresponding single-orbital Hubbard model when considering the application to DMFT, but also show a DMFT calculation for a two band Hubbard model.

For the single-impurity Anderson model
the basic object of study is the imaginary time action
$S=-\int d\tau d\tau' \sum_{\sigma} c_{\sigma}^{\dagger}(\tau)
 [ G_0^{-1}(\tau\!\!-\!\!\tau') + \mu \delta(\tau\!\! -\!\! \tau') ]c_{\sigma}(\tau')
+ U\int d\tau c_{\uparrow}^{\dagger}(\tau) c_{\downarrow}^{\dagger}(\tau)c_{\downarrow}(\tau)c_{\uparrow}(\tau)$,
%
%
with spin $\sigma=\uparrow,\downarrow$. The non-interacting impurity Green's function $G_0$ describes the correlations induced by the coupling to the surrounding non-interacting bath. In complex frequency space, $G_0(z)$ is an analytic function with poles (branch cut) on the real frequency axis. The task is to calculate the Green's function $G_{\sigma}(\tau-\tau')=-\langle Tc_{\sigma}(\tau)c_{\sigma}^{\dagger}(\tau')\rangle_S$. Subsequently we will assume no magnetic order and drop the spin index and instead of the Green's function we can consider the self energy $\Sigma$ given by $G^{-1}(z)=G_0^{-1}(z)+\mu-\Sigma(z)$.  
For the quantum impurity problem, $G_0$ corresponds to the bare (non-interacting) density of states which we take to be semicircular $\rho_0(\omega)=(2/\pi)\sqrt{1-\omega^2}$ and let the half-bandwidth be our unit of energy. 

Consider a representation of $G_0$ in terms of a large number of poles on the real axis $G_0(z)=\sum_{i=1}^M\frac{\tilde{a}_i}{z-b_i}$ with $\sum{\tilde{a}_i}=1$, which for the purpose of doing numerical calculations can be very good for a large number ($10^4$ or more) of poles. Assume that the poles and residues are such  that they can be grouped in $N$ smaller groups of size $n$ (the total number of poles is thus $M=Nn$) such that the total residue in each group is $1/N$. (There are many inequivalent ways of grouping the poles, we make an unbiased choice, grouping the poles randomly.) 
 We rewrite 
\begin{equation}
\label{g0split}
G_0(z)=\frac{1}{N}\sum_{\nu=1}^{N}\sum_{j=1}^{n}\frac{a^{\nu}_j}{z-b^{\nu}_j}=\frac{1}{N}\sum_{\nu}G^{\nu}_0(z)\,,
\end{equation}
where the residues are renormalized by a factor $N$ such that 
the Green's functions $G^{\nu}_0(z)$ are properly normalized for an n-level system, $\sum_j a^{\nu}_j=1$.

\begin{figure}
\includegraphics[scale=1]{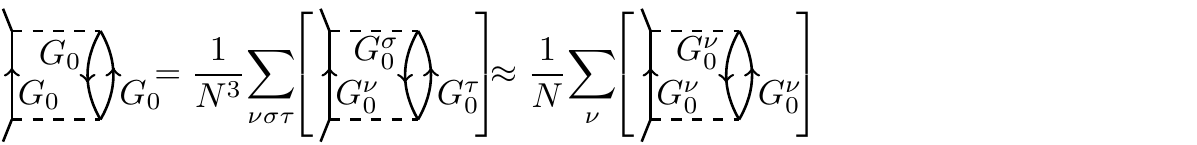}
\caption{\label{approxfig} Basic approximation of the formalism corresponding to grouping of poles (Eq.~\ref{g0split}) and ignoring cross correlations between groups, illustrated here for a 2nd order diagram.
\\[-6mm]
}
\end{figure}

The self energy is given by all one particle irreducible (1PI) diagrams in terms of the four point vertex $U$ and the two point vertex $-\mu$ connected by $G_0$.~\cite{Negele:1998aa}
For every diagram, we make the following approximation exemplified in Fig.~\ref{approxfig} by a 2nd order diagram (omitting $\mu$ insertions).

The approximation is thus to replace cross correlations between different Green's functions by internal correlations, giving $\Sigma\approx\frac{1}{N}\sum_{\nu}\Sigma^{\nu}$
where $\Sigma^{\nu}$ is the self energy related to $G_0^{\nu}$.

Considering also the chemical potential $\mu$ on the interacting site, all 1PI diagrams include the diagram with the vertex $-\mu$ by itself as well as all insertions of $-\mu$ into the 1PI diagrams constructed from the vertex $U$. Within the same approximation, $\mu$ acts as a chemical potential on each subsystem such that  $G_0(z)+G_0^2(z)(-\mu)+...\approx \frac{1}{N}\sum_{\nu}\frac{1}{(G_0^{\nu}(z))^{-1}+\mu}$ and we find that the approximation corresponds to
the expression
\begin{equation}
\Sigma-\mu\approx\frac{1}{N}\sum_{\nu=1}^N(\Sigma^{\nu}-\mu)\,,
\end{equation}
which is the basis of the present formalism. 

Importantly, $\Sigma^{\nu}$ contains all 1PI diagrams of $G_0^\nu$, it is the exact self energy of the quantum impurity action
$S$, 
with $G_0$ replaced by $G_0^{\nu}$, a problem that can be 
mapped to an Anderson impurity model with a single interacting site coupled to $n-1$ bath levels. The Anderson model is formulated in terms of a Hamiltonian which can be diagonalized numerically for small $n$ and the self energy calculated as 
\begin{equation}
\label{dyson}
\Sigma^{\nu}(z)-\mu=(G_0^{\nu}(z))^{-1}-(G^{\nu}(z))^{-1}\,.
\end{equation}
Note that it is not an option to work with the sample averaged Green's functions 
instead of the self energy; 
$\frac{1}{N}\sum_\nu G^{\nu}$ 
and
$\frac{1}{N}\sum_\nu G_0^{\nu}$
is not a proper pair of interacting and non-interacting Green's functions, 
they do not obey $G_0(z)=0\Rightarrow G(z)=0$, 
as required by the Dyson equation.
%

Now, $G_0^{\nu}=\sum_{j=1}^{n}\frac{a^{\nu}_j}{z-b^{\nu}_j}$ is mapped to the Green's function $G_0^{\nu}=1/(z-\epsilon_0^{\nu}-\sum_{j=1}^{n-1}\frac{(V^{\nu}_j)^2}{z-\epsilon^{\nu}_j})$ of the Anderson model
$H_0=\epsilon_0^{\nu} \sum_{\sigma}c_{\sigma}^{\dagger}c_{\sigma}
+\sum_{\sigma,j=1}^{n-1}[V_j^{\nu}(c_{\sigma}^{\dagger}c_{j\sigma}+h.c.)+\epsilon_j^{\nu}c_{j\sigma}^{\dagger}c_{j\sigma}]$
by solving for the parameters $\epsilon_i$ and $V_i$ according to $\epsilon_i^{\nu}:G_0^{\nu}(\omega=\epsilon_i^{\nu})=0$, $\frac{dG_0^{\nu}}{d\omega}|_{\epsilon_i}=-1/(V_i^{\nu})^2$,
and $\epsilon_0^{\nu}=-\sum_j a^{\nu}_j b^{\nu}_j$.
The full Hamiltonian is $H=H_0-\mu\sum_{\sigma} c_{\sigma}^{\dagger}c_{\sigma}
+ Uc_{\uparrow}^{\dagger}c_{\downarrow}^{\dagger}c_{\downarrow}c_{\uparrow}$ and the corresponding Green's function $G^\nu(z)$ given by the Lehmann representation by summing over the complete set of eigenstates.~\cite{Fetter:2003aa}

The $n$-level systems is derived from a representation of the full $G_0$ in terms of a large number of poles grouped into sets of poles with normalized residues, Eq. \ref{g0split}. An exact representation of $G_0(z)$ is a distribution of poles given by the spectral function $A_0(\omega)=-\frac{1}{\pi} \textrm{Im}[G_0(w+i0^+)]$ and in practice we will use this as a probability distribution for generating sequences of $n$ random pole locations with equal residues. (Other sampling procedures are conceivable.) A large ensemble of such groups will give a good representation of $G_0$.
However, one show that the approximation (illustrated in Fig.~\ref{approxfig}) overestimates the contribution of low energy spectral weight in $G_0$. We have found that this can be compensated by imposing the constraint that the ground state particle number of the interacting and non-interacting $n$-level systems coincide, discarding configurations that fail to satisfy this criterion. (This also means that the sampling will not be a completely faithful, as seen in Fig. \ref{QMC_comp}.)

\begin{figure}[t]
\includegraphics[scale=1]{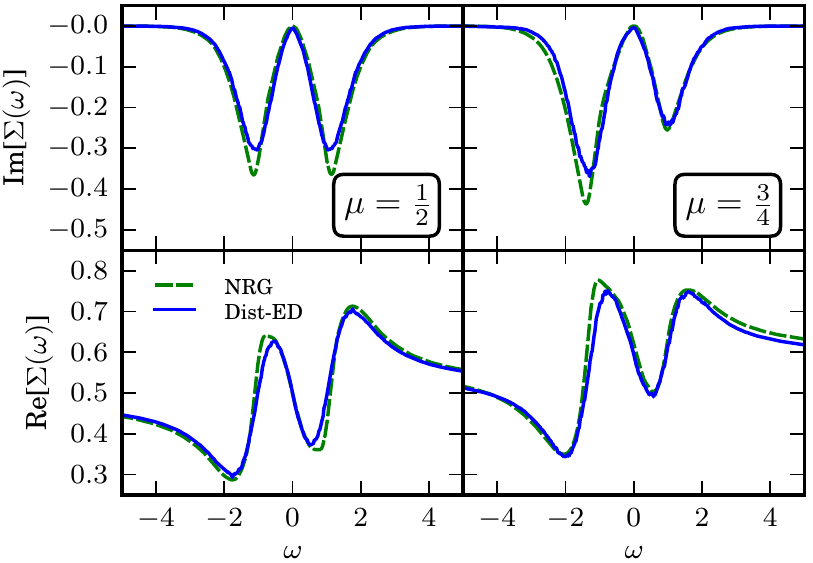}
\caption{\label{GreteCF}(Color online) DMFT real frequency self-energy $\Sigma(\omega)$ (solid line) for U=1, T=0, at and away from half-filling, $\mu=0.5$ and $\mu=0.75$ respectively, corresponding NRG results (dashed line) from Grete {\em et al.}, Ref.~\onlinecite{Grete:2011yq}. Model size $n=6$ and $2 \cdot 10^4$ samples with acceptance ratios $89\%$ and $81\%$.
\\[-6mm]
}
\end{figure}

The suggested operational procedure for a particle-hole symmetric system is:  \\
$\bullet$ Use $A_0(\omega)$ as a probability distribution for generating $N$ sets of $n$-poles, with normalized residues, giving the Green's functions $G_0^{\nu}$, $\nu=1,...,N$.\\
$\bullet$ 
For each $\nu$ check that the ground state of the interacting Hamiltonian has the same particle number as the non-interacting, else discard the configuration $\nu$. Calculate the self energy $\Sigma^{\nu}$ by mapping the problem to a Hamiltonian and using Eq.~\ref{dyson}.\\
$\bullet$ Add up the self energy $\Sigma=\frac{1}{N}\sum_{\nu}\Sigma^{\nu}$ 

When applied to DMFT $G_0$ is calculated from the DMFT equations for the local Green's function $G_L$
$G_L(z)=\int \frac{d\omega'}{2\pi}\frac{\rho_0(\omega')}{z-w'+\mu-\Sigma(z)}$ and 
$G_{0}^{-1}(z)=G_L^{-1}(z)+\Sigma(z)-\mu\,$
which follows from integrating out all degrees of freedom except those of one single (impurity) site. Given $G_0$ the impurity problem is solved for the self-energy $\Sigma$ which gives a closed set of equations that are solved self-consistently.

Away from half filling the standard formalism is poorly suited for the sampling, as the particle number of $G_0$ and $G_L$ may be very different.
We will use a formally equivalent expression, defining $\tilde{\Sigma}(z)=\Sigma(z)-\Sigma_0$ where $\Sigma_0$ is a real constant. 
In terms of this we write 
\begin{equation}
G_L(z)=\left[(\tilde{G}_0(z))^{-1}-\tilde{\Sigma}(z) \right]^{-1}
\end{equation}
where $\tilde{G}_0^{-1}(z)=G_0^{-1}(z)+\mu-\Sigma_0$ and 
choose $\Sigma_0$ such that 
$\int d\omega \, \textrm{Im}[\tilde{G_0}(\omega)]n_F(\omega)=\int d\omega \, \textrm{Im}[G_L(\omega)] n_F(\omega)$, the two Green's functions give the same occupation. Thus starting each step of the DMFT iteration we would solve for $\Sigma_0$ by fitting the particle numbers,  use $\Sigma_0$ as the effective chemical potential for the $n$-level systems and the spectral function of $\tilde{G_0}$ as probability distribution. (For half filling $\Sigma_0=\mu=u/2$ and this step is trivial.) For the quantum impurity model we work with $G=1/((G_{0,bare})^{-1}+\mu-\Sigma)$ instead of $G_L$ and choose $\Sigma_0$ such that $G$ and $\tilde{G_0}=1/(G_0^{-1}+\mu-\Sigma_0)$ have the same particle number. Here we need to converge $\Sigma_0$, because $\Sigma$ enters into $G$ and the particle numbers of $G$ and $\tilde{G}$ should be same for the best sampling. In what follows, we present several different calculations and compare to NGR and CT-QMC.

\begin{figure}[t]
\includegraphics[scale=1]{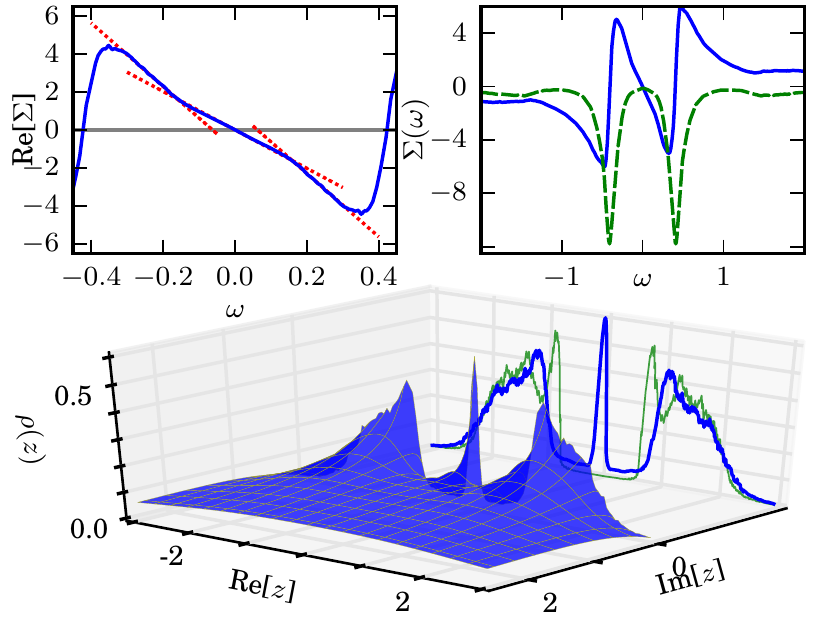}
\caption{\label{DMFT_fig}
(Color online) DMFT calculation for $U=2.8$, $T=0$, (upper left panel) real part of $\Sigma(\omega)-U/2$ (solid line) with kinks indicated fitted dotted lines, (upper right panel) full view of the Self-energy.
(lower panel) $-\textrm{Im}[G(z)]$ in the upper complex plane (grid-surface) and on the real axis (solid line), coexisting insulating solution (thin solid line). Model size $n=6$ and $3 \cdot 10^4$ samples with $49\%$ acceptance ratio.
\\[-6mm]
}
\end{figure}

\begin{figure}[b]
\includegraphics[scale=1]{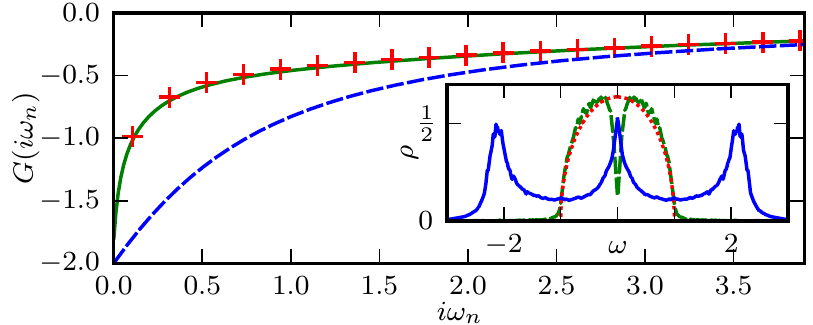}
\caption{\label{QMC_comp}(Color online) Anderson impurity at half-filling, $U=3$ and $\beta=30$.
%
%
%
Interacting Green's function $G(z)$ on the imaginary axis (solid line) compared to CT-QMC results (crosses) using the Triqs-code,\cite{triqs_project, Werner:2006rt, Werner:2006qy, Boehnke:2011fk} bath Green's function $G_0$ (dashed line). (inset) $\rho(\omega)=-\textrm{Im}[1/\pi G(\omega + i\delta)]$ (solid line), bath DOS $\rho_0(\omega)$ (dotted line), and sampled $\rho_0$ discarding configurations (dashed line).
Model size $n=6$ and $10^4$ samples with $65\%$ acceptance ratio.
\\[-0.6cm]
}
\end{figure}

In Figure \ref{GreteCF} we show DMFT calculations for $U=1$, $T=0$, with $\mu=0.5$ (half filling) and $\mu=0.75$ to compare with published real frequency NRG~\cite{Grete:2011yq} results for the self energy. For the system away from half filling we use a sampling where the particle number of the non-interacting samples are free to vary between $0$ and $2n$ (in units of 2) guided by the probability distribution. The results are in  strikingly good agreement considering that there are no free parameters in the formalism.  


In Figure \ref{DMFT_fig} we show results for $U=2.8$, close to the $T=0$ Mott transition at $U\approx 3$,\cite{Bulla:2008kx} both the metallic and insulating solutions. Clearly, the method is sophisticated enough to capture fine structure in the self energy, corresponding to kinks in the quasiparticle dispersion. \cite{Byczuk:2007kx, Grete:2011yq}   

Figure \ref{QMC_comp} is a finite temperature calculation at $\beta=30$ ($T=1/30$) for the Anderson impurity model at $U=3$. We find good agreement with CT-QMC
calculations for the impurity Green's function $G(z)=1/(G_{0,bare}^{-1}(z)+\mu-\Sigma(z))$ on the Matsubara frequencies $z=i\omega_n=i\frac{\pi}{\beta}(2n+1)$, with maximum error for $n=6$ is 4\%. 
Interestingly, the real frequency density of states is quite different from that deduced from MAXENT. \cite{Hafermann:2009aa} The central (Kondo) peak is similar but the Hubbard bands are much more distinct.

\begin{figure}
\includegraphics[scale=1]{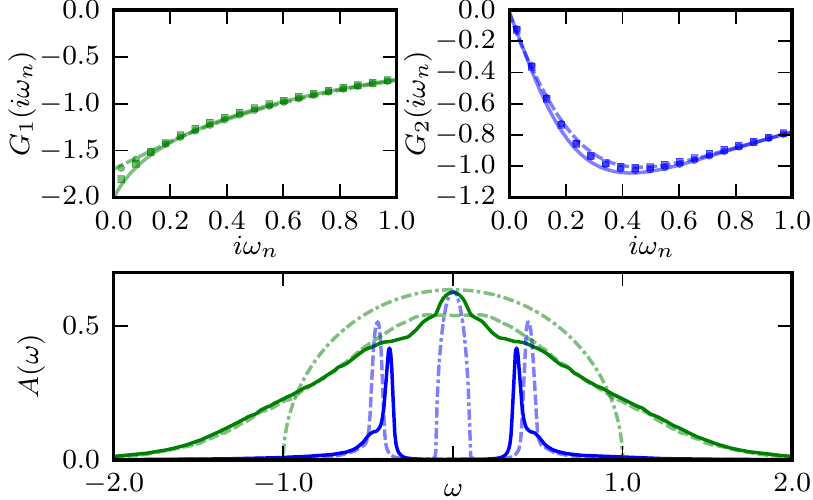}
\caption{\label{twoband_fig}(Color online) 
Two band Hubbard model at $T=0$ with $U = 1$, $J = 0.2$, and half-bandwidths $W_1=1$ and $W_2=0.1$ in the orbital selective Mott phase. 
With density-density (dashed lines) and full Hund's (solid lines) interaction, the former is a non-Fermi liquid with $\textrm{Im}[\Sigma_{1}(0)]\neq 0$. 
(Top) $\textrm{Im}[G(z)]$ along the imaginary axis compared to CT-QMC\cite{triqs_project, Werner:2006rt, Werner:2006qy, Boehnke:2011fk} ($\beta=120$) for the wide (left) and narrow (right) band.
(Bottom) Real-frequency spectral functions (narrow band scaled by $1/10$), non-interacting case (dash-dotted lines) density-density (dashed lines), and Hund's (solid lines) interaction.
Model size $n=6$ and $\approx 10^4$ accepted samples.
\\[-0.6cm]
}
\end{figure}

To show the generality of the method we also present a calculation for the two band Hubbard model both with density-density and full Hund's rule coupling, for model details see 
Refs.~\onlinecite{Liebsch:2006aa, Medici}. The spectral functions for the bands in the orbitally selective Mott phase are shown in Fig.~\ref{twoband_fig} and the imaginary frequency Green's functions are compared to CT-QMC results. The sampled systems have a total size of $n=6$ with three (one) bath sites for the metallic (insulating) band. For the system with only density-density interactions, which is a non-Fermi liquid, we have used the fact that the spin is conserved in each orbital, to calculate two independent self-energies for the wide band, depending on the spin in the narrow band, see Biermann {\em et al}. Ref.~\onlinecite{Medici}.

In summary, we have presented a formalism for calculating the full analytic self energy of quantum impurity models by using a representative distribution of exactly solvable Anderson impurity models. The method is simple to implement and the initial studies shows that the method can give very good results. The calculations in this paper were done a single desktop computer over time periods of 10-40 hours, but the formalism is well suited for parallel computing which will be the key to considering larger $n$ models. A natural extension is to apply the formalism to more general models, including multi orbital and cluster generalizations of DMFT and to calculate other dynamical correlations. 

We want to thank, Stellan \"Ostlund for valuable discussions and the use of his Mathematica routines for fermions, Ansgar Liebsch for valuable discussions on low frequency features in the self-energy, and Sebastian Schmitt for providing the NRG results. The work was supported by the Swedish Research Council (grant no.\ 2008-4242) and the Mathematics - Physics Platform ($\mathcal{MP}^{\textsf{2}}$) at the University of Gothenburg.


\end{document}